# Knowledge AI: New Medical AI Solution for Medical image Diagnosis

**Yingni Wang[1], Shuge Lei[1,2], Jian Dai[1], Kehong Yuan[1*]**

[1] Graduate School at Shenzhen, Tsinghua University, Shenzhen, China
[2] Computer Science and Engineering, University of South Carolina, SC, United States

Corresponding author: Kehong Yuan (yuankh@sz.tsinghua.edu.cn)

**ABSTRACT** The implementation of medical AI has always been a problem. The effect of traditional perceptual AI algorithm in medical image processing needs to be improved. Here we propose a method of knowledge AI, which is a combination of perceptual AI and clinical knowledge and experience. Based on this method, the geometric information mining of medical images can represent the experience and information and evaluate the quality of medical images.

1. **Introduction**

There is a big difference between natural images and medical images. Natural images are generally obtained by visible light in life, such as landscape and people photos. In addition, data obtained by lidar or structured light also fall into the category of natural images. Natural image is the data representation of visual information transmitted by nature. On the other hand, medical images are mostly derived from human tissues, which are the results of imaging from certain parts in vivo or in vitro. The main imaging methods include optical microscopy, ultrasound, MRI, CT and PET. People can recognize natural images accurately through continuous cognition and learning, but medical images can only be recognized through long-term training and combined with clinical experience.

Deep learning has achieved very good results in natural image recognition, segmentation and target detection, and has good generalization performance. In 2016, ResNet won the champion at ILSVRC （ImageNet Large Scale Visual Recognition Challenge） with 96.4% accuracy, which is far better than traditional machine learning algorithms. Common natural images are very familiar to us, and they are easy for human beings to understand and accept. At the same time, the black box problem in the natural image AI algorithm can usually be ignored, and there is no particularly high requirement for the interpretability of the algorithm. But for medical AI, we pay more attention to the interpretability of results, whose realization must rely on clinical knowledge and experience.

The current AI is in a state of weak artificial intelligence, which only has the ability of perception and comparison, and lacks the learning, understanding and reasoning of information, which is called perceptual AI. The existing perceptual AI methods cannot fully reflect the knowledge and experience information hidden in medical images. However, the performance of deep learning in the field of medical image processing is not satisfactory. Esteva et.al used GoogleNet based on Inception V3 to classify skin images into three categories of benign, malignant and non-neoplastic lesions, but the accuracy of CNN could only be reached 72%[1]. This means that three out of every 10 patients will be misdiagnosed, a figure that is clinically unacceptable.

Knowledge AI, on the other hand, aims to give artificial intelligence abilities similar to those of the human brain, that is, to grasp knowledge and reason. We embed the geometric registration method and digital twinning method into the existing perceptual AI model to mine the shape and structure information of the detected parts, and make comparison through the standard map, so as to realize the expression of knowledge and experience. With natural images, we can deepen our understanding through constant contact. But this is not set up for the medical images, we cannot achieve qualitative breakthrough by the accumulation of quantity. That is to say, we cannot make ourselves capable of being a doctor by viewing medical images many times, because doctors' diagnostic ability of medical images is acquired through accumulated practice on the basis of their rich experience. For AI algorithms, perceptual AI can have a deeper understanding of images through repeated learning and training, but merely improving the ability of understanding cannot help perceptual AI to have the function of independent analysis and diagnosis.

Compared with natural images, medical images contain abundant disease information, but the recognition of medical images requires rich clinical knowledge and experience. This also determines that medical AI is not simply perceptual AI(P-AI), but a combination of perceptual AI and clinical knowledge and experience, which called knowledge AI(K-AI),as shown in Fig.1.

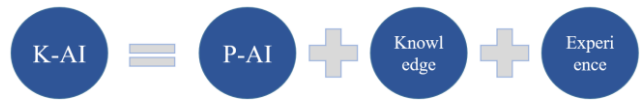

Fig.1 The architecture of K-AI.

## 2. METHOD

### 2.1 PIPLINE of KNOWLEDGE AI

#### 2.1.1 Perceptual AI

Perceptual AI uses an attention mechanism to locate parts of an image where key information is hidden, learning the importance of different local features. Scientific studies have shown that when human beings observe the environment, their brains tend to focus only on a few important local regions to obtain the information needed to construct a description of the object and the surrounding environment. The attention mechanism of deep learning is exactly the same. The neural network learns the importance of different parts and combines them together. The perceptual AI based on attention mechanism can improve the efficiency and accuracy of algorithm operation. However, the traditional deep learning method uses a series of convolutional neural networks to automatically extract features from all regions of the image, which consumes a lot of time and computing power. The core of perceptual AI is the attention mechanism, which is summed up by the habit of observing the environment.

The attention mechanism can be regarded as the use of information in the feature mapping of the later layers of the network to select and locate the most discriminating part of the input signal. The attention module is a branch network with a feature map node in the main network. Generally speaking, the attention mechanism of the feature map response of different layers in the network is different. According to experience, in the shallower layer network, the attention characteristic map focuses on the background and other areas, while in the deep layer structure, it focuses on the objects to be classified. Furthermore, it shows that the deep level feature map has higher abstractness and semantic expression ability[2].

#### 2.1.2 Knowledge

Different from natural images, medical images have certain regularity in terms of spatial structure and content, and the content in the image is relatively certain. If you take an ultrasound of the heart, for example, the atria and the ventricles must be spatially adjacent to each other and surrounded by the epicardium.

Therefore, in the medical AI algorithm, we can add certain prior knowledge to the model, which will greatly improve the performance of the algorithm. In this model, geometric registration and knowledge map are added to the perceptual AI mentioned in the previous section to mine the information in the image, and anatomical knowledge, organ structure, imaging features and other prior knowledge are added to the model. For example, in the analysis of breast ultrasound images, the neural network will look for the connection between local tissues and surrounding parts as well as other tissues in previous and next frame images, so as to better capture the characteristics of organ structure. At the same time, we also used geometric registration method to construct the standard brain atlas construction model based on brain MRI images. The construction of standard brain atlas plays an important role in recognizing the complexity of brain structure. For a fixed input, we get the Jacobian determinant (JD) and curl vector (CV) by geometric method and use them as the input of the neural network to get the standard brain atlas by training and geometric registration method.

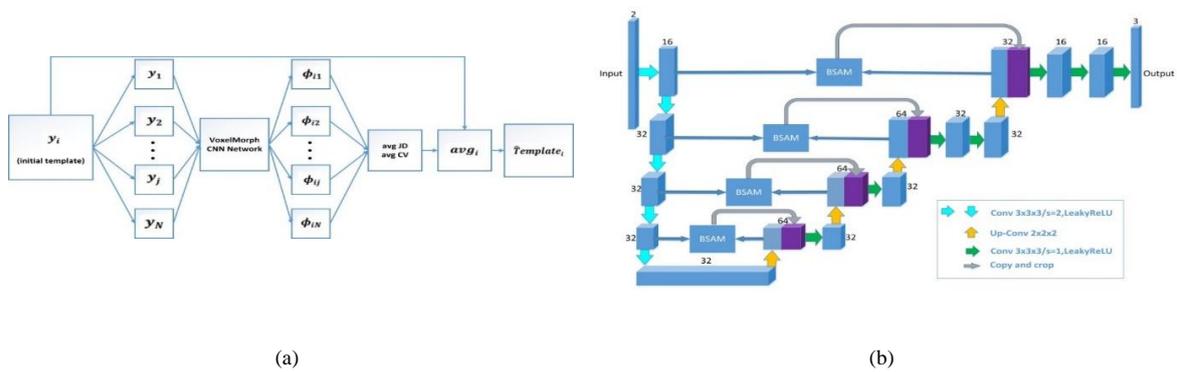

(a) (b)

Fig.2 The standard brain atlas construction model:(a) Standard atlas generation method,(b)The structure of Voxel-Morph CNN Network.

### 2.1.3 Experience

The diagnosis of clinical diseases not only requires medical knowledge, but also depends on the clinical experience accumulated by doctors in practice. Different doctors may perceive the same ultrasound image differently, and less experienced doctors may not be able to capture key information about the diagnosis of the disease. In this study, geometric methods are used to conduct information mining on medical images, so as to obtain deeper diagnostic information, which can help inexperienced young doctors make correct diagnosis and help them carry out their work with higher accuracy and confidence. The expression of experience is mainly embodied in two aspects, one is the Quality control of ultrasonic image, the other is the explicit expression of texture features in medical images.

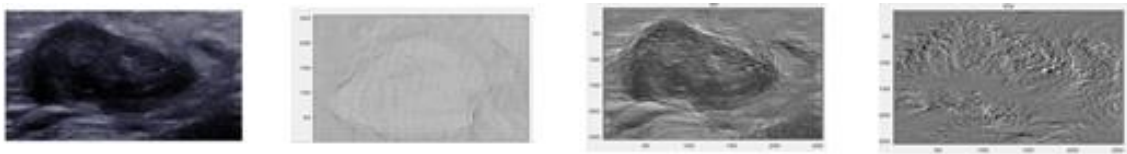

Fig. 3. Different doctors have different perceptions of ultrasound images:(a)Breast cancer lesion in ultrasound image(b) The grid image from (a),(c) Image generated by JD from (a),(d)(e)(f) Image generated by CV from (a).

It is very important to accurately locate the standard incisions from the dynamic ultrasound images for subsequent physiological parameters measurement and disease diagnosis when using ultrasonic images for disease diagnosis. In the past, standard section identification was mainly made by doctors manually, but this method not only needs a lot of time and energy, but also depends on the clinical work experience of sonographers[3]. The automatic identification algorithm of ultrasonic standard section based on deep convolutional network proposed in this paper can quickly and accurately locate the standard section in the process of physician scanning, thus helping young doctors improve the efficiency of ultrasonic diagnosis and relieve working pressure.

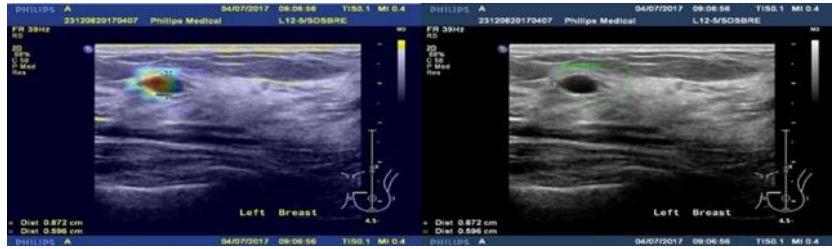

Fig. 4.   Standard section of ultrasound images and diagnosis of lesions

The cause of cancer is that normal cells grow out of control under the combined action of genetic, environmental and other factors, and then develop into a kind of invasive disease. In the process of disease development, the variation of cell genome will accumulate gradually, and the cell subsets with different biological behaviors will be differentiated in different directions. At the macroscopic level, the tumors were heterogeneous with differences in cell phenotypes within the tumors, and the imaging results showed differences in texture features. Studies have shown that tumor heterogeneity is closely related to the treatment regimens and the prognostic effect. However, the distinguish of tumor heterogeneity requires rich clinical experience, and inexperienced young doctors are easy to ignore the texture that is not obvious in the image[4].

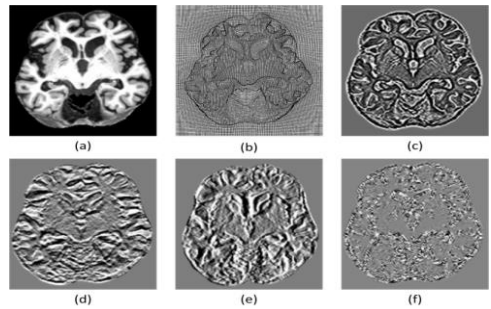

Fig. 5.   Representation and AI realization of physician experience :(a)  T1 MRI images,(b)  The grid image from (a),(c)  Image generated by JD from (a),(d)(e)(f)  Image generated by CV from (a), 3 channels.

Different from CT and MRI medical imaging, the quality of ultrasound image is greatly affected by the physician's operation. In general, high quality ultrasound images have better diagnostic results. We propose an objective evaluation model of ultrasonic image quality based on meta learning to comprehensively evaluate the acquired ultrasonic image.

For the same subject, the images collected by the same device with different imaging parameters are different. Take MRI imaging as an example, the image resolution will change significantly under 1.5T and 3.0T magnetic field intensity. In order to avoid the impact of image differences caused by different equipment and parameters on the final analysis results, we proposed the SRGAN network model to improve the robustness of image acquisition.

**2.1.4 Combination of perception, knowledge and experience**
1) The serial model：Firstly, the perceptual AI algorithm is used to analyze the medical data, and then the prior knowledge and clinical experience are combined for analysis and judgment.
2) The coupled model: It can directly act on image acquisition and guide sonographers to obtain high quality images. We first process the input through the perceptual AI structure, and then evaluate the network output by embedding knowledge and experience information through geometric methods and digital twinning techniques. When the output is inconsistent with the expectation, the network structure is adjusted by the penalty function; otherwise, the current output is taken as the final result.

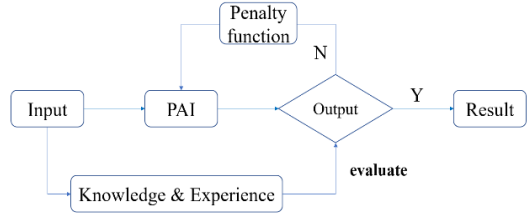

Fig. 6. The coupled method we proposed

3) The parallel model: It is a results-oriented method to evaluate medical images at the same time of data acquisition, which does not have a guiding role.

## 2.2 DATA AND PREPROCESSING

Obtaining medical data has always been a difficult task, and there are no suitable public data set resources on the Internet. The data set of our experiment comes from the database of the Ultrasound Imaging Department of Peking University Shenzhen Hospital. We collected 6,860 pictures, including 2,065 images with benign lesions and 3,495 images with malignant lesions, and 1,300 images without lesions.

Table 1 The dataset used in our experiment

| Dataset | Images with benign nodules | Images with malignant nodules | Images without nodules |
| --- | --- | --- | --- |
| Training Dataset | 1870 | 3160 | 0 |
| Validation Dataset | 195 | 335 | 1300 |

In the data marking stage, five doctors used the open-source software LabelImg to label these data sets, which were collected from 2804 patients. In order to ensure the accuracy of the data, one doctor marks the image while the other doctor checks it.

The target detection model must have a target during the training process; only the image of the lesion is helpful to the training model. Therefore, as shown in Table 1, we finally selected 5030 lesion images as the training set, 530 lesion images, and 1300 disease-free images as the validation set.

## 2.3 METRICS IN OUR EXPERIMENT

The purpose of the experiment is to improve the clarity of ultrasound images and improve the detection accuracy of the target detection model. Therefore, the evaluation indicators we use in our experiments are mainly from the field of target detection and digital imagery.

Table 2 The metrics used in the experiments

| Evaluation index | Method of calculation |
| --- | --- |
| Precision | $\frac{TP}{TP + FP}$ |
| Recall | $\frac{TP}{TP + FN}$ |
| Specificity | $\frac{TN}{TN + FP}$ |
| mAP | Area covered under PR curve |
| MSE | $MSE = \frac{1}{H \times W} \sum_{i=1}^{H} \sum_{j=1}^{W} (X(i,j) - Y(i,j))^2$ |
| PSNR | $PSNR = 10 \cdot \log_{10} \left(\frac{MAX_I^2}{MSE}\right)$ |
| SSIM | $SSIM(x,y) = \frac{(2\mu_x \mu_y + c_1)(2\sigma_{xy} + c_2)}{(\mu_x^2 + \mu_y^2 + c_1)(\sigma_x^2 + \sigma_y^2 + c_2)}$ |
| RTP | The task reduced by AI machine |

As shown in Table 2, in order to evaluate the effect of lesion feature detection, we used four indicators: precision, recall, sensitivity, and mAP value. In the calculation formula, TP and FN represent the number of positive samples inferred as positive and negative samples, and FP and TN represent the number of negative samples inferred as positive and negative. Obviously, Precision reflects the model's ability to accurately identify lesion images, and Specificity reflects the model's ability to accurately identify disease-free images. Draw a curve composed of recall score and precision score on the coordinate axis. The area of the curve represents the AP value, which is usually used to comprehensively measure the accuracy of the detector model. The classification task is not the same as the target detection task. It only makes a judgment on whether it is diseased or not, rather than the location of the lesion in the ultrasound image. Therefore, our experiment is divided into two tasks: lesion detection and image classification, but the scoring principle is the same.

We have also defined an RTP indicator (Reduced Task Proportion), which means that the AI-assisted diagnostic device helps doctors reduce the workload. It should be pointed out that our Multi-AI Model does not make a definite evaluation for every input picture. Only when the evaluations of all sub-models are consistent, the system will give its judgment, and all other pictures will be given to the doctor. Since it only selects certain pictures to make judgments, we use RTP indicators to comprehensively evaluate the practical significance of the model.

## 3. Evaluation methodology

### 3.1 Implementation of geometric methods

We use geometric methods to mine the experiential information in brain MRI images, and represent the experiential information through AI models. The method based on geometric topology can explore the structure, deformation and other information in the image in a deeper level, so as to help doctors find the critical diagnostic information for the disease but difficult to find due to lack of experience.

### 3.2 Quality control model

#### 3.2.1 Standard section recognition model of ultrasonic image

In this study, a model of ultrasonic standard section recognition was built based on deep convolutional neural network. We realize standard section recognition mainly by looking for the key organs which play an important role in locating standard section and their spatial position relations.Different from the traditional classification method based on convolutional neural network, adding prior knowledge such as spatial location and structure can improve the learning and reasoning ability of the network.

#### 3.2.2 SRGAN network quality control model

The GAN network consists of a generator model (G) and a discriminative model (D). The G model is responsible for generating data that is as close to the real sample as possible, and the D model is used to score the output result, which is to judge whether the generated sample is true or false. As shown in Fig.8, the input of the G model of the SRGAN network is a low-resolution image $I^{LR}$, and the high-resolution output image $I^{SR}$ will be used as the input of the D model together with the original image $I^{HR}$.

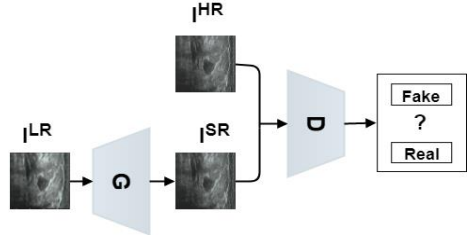

Fig. 7.  The framework of SRGAN

When training SRGAN, it uses a four times downsampling factor to obtain low-resolution images and optimizes in an alternating manner to solve the minimum-maximum adversarial problem:

$$\min_{\theta_C}\max_{\theta_D} \mathbb{E}_{I^{HR}_R \sim p_{\text{train}}(I^{HR})}[\log D_{\theta_D}(I^{HR})] + \mathbb{E}_{I^{LR} \sim p_C(I^{LR})}\left[\log\left(1 - D_{\theta_D}\left(G_{\theta_G}(I^{LR})\right)\right)\right] \quad (1)$$

Different from previous works, they defined a novel perceptual loss using high-level feature maps of the VGG network[5], combined with a discriminator that encourages solutions perceptually hard to distinguish from the HR reference images[6]. The VGG loss is defined as the Euclidean distance between the feature representation of the reconstructed image $G_{\theta_G}(I^{LR})$ and the reference image $I^{HR}$:

$$l^{SR}_{VGG/i.j} = \frac{1}{W_{i,j}H_{i,j}}\sum_{x=1}^{W_{i,j}}\sum_{y=1}^{H_{i,j}} -(\phi_{i,j}(I^{HR})_{x,y} - \phi_{i,j}\left(G_{\theta_G}(I^{LR})\right)_{x,y})^2 \quad (2)$$

The method of independent alternating iterative training makes the two networks oppose to each other. In the end, the goal of the G network can "deceive" the D network, and the G network is what we need[7]

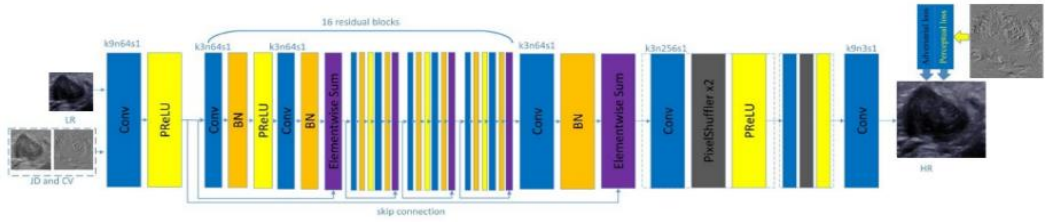

Fig. 8.  Our proposed architecture of Generator Network with corresponding kernel size (k), number of feature maps (n) and stride (s) indicated for each convolutional layer.

The loss function of SRGAN network is defined as follows:

$$l^{SR}_{CV} = \frac{1}{W_{i,j}H_{i,j}}\sum_{x=1}^{W_{i,j}}\sum_{y=1}^{H_{i,j}}\left(CV(I^{HR})_{x,y} - CV\left(G(y, I^{LR})\right)_{x,y}\right)^2 \quad (3)$$

Here $W_{i,j}$ and $H_{i,j}$ describes the dimensions of the CV feature maps from $I^{HR}$ and $G(y, I^{LR})$.

### 3.2.3 Objective Evaluation Model of Ultrasonic Image Quality Based on Meta-learning

In order to evaluate the quality of acquired images with objective criteria, an objective evaluation model of ultrasonic image quality based on meta-learning was established. The model firstly refines the evaluation tasks of different clinicians on the quality of ultrasound images into the selection of meta-tasks, which is achieved by measuring the similarity of the gradient direction between tasks and sorting them by algorithm.

After determining the meta-tasks, the ultrasonic image quality evaluation model based on the joint gradient optimization was constructed. Firstly, the meta-tasks were sampled into the meta-training set, and the joint gradient was optimized through the deep regression network to construct the quality evaluation model.

The mathematical subjective evaluation method of ordered attributes is as follows:

$$(x_1, x_2, \ldots, x_k) = \begin{pmatrix} d_{11}, d_{12}, \ldots, d_{1k} \\ \ldots, \ldots, \ldots \ldots \ldots \\ d_{k1}, d_{k2}, \ldots, d_{kk} \end{pmatrix} \begin{pmatrix} p_1 \\ \ldots \\ p_k \end{pmatrix} \quad (4)$$

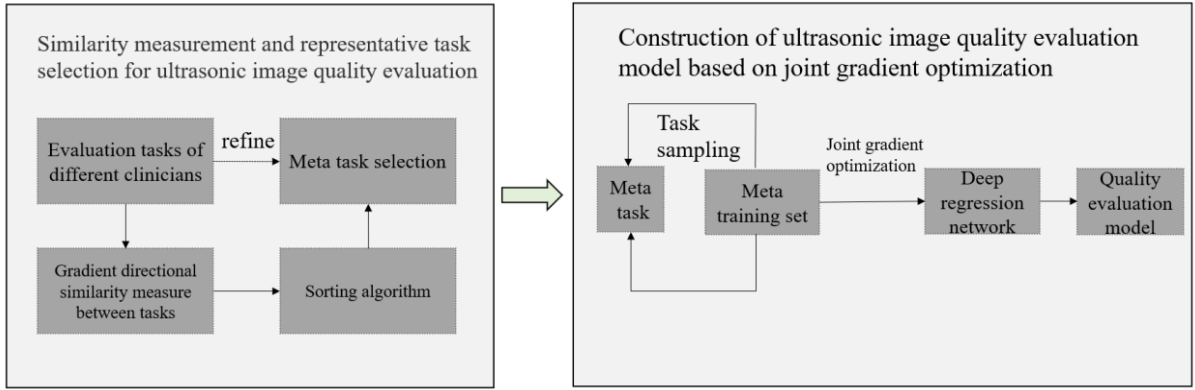

Fig. 8. Our proposed architecture of quality evaluation model.

## 4. Conclusion

Medical image recognition must rely on knowledge and experience because of the particularity of its imaging and content. By combining doctors' clinical knowledge and experience with traditional perceptual AI, the knowledge AI proposed by us can endow algorithm with the ability to infer and think, fully mine the information in medical images, and obtain better performance. It has a broad application prospect in the future medical industry.